\documentstyle[12pt]{article}

\renewenvironment{thebibliography}[1]
{\normalsize
 \begin{list}{[\arabic{enumi}]}
 {\usecounter{enumi} \setlength{\parsep}{0pt}
  \setlength{\itemsep}{3pt} \settowidth{\labelwidth}{[#1]}
  \sloppy}}
{\end{list}}

\newcommand{\bea}{\begin{eqnarray}}
\newcommand{\eea}{\end{eqnarray}}
\newcommand{\beq}{\begin{equation}}
\newcommand{\eeq}{\end{equation}}
\newcommand{\simgt}{\hbox{ \raise3pt\hbox to 0pt{$>$}\raise-3pt\hbox{$\sim$} }}
\newcommand{\simlt}{\hbox{ \raise3pt\hbox to 0pt{$<$}\raise-3pt\hbox{$\sim$} }}

\def\lsa{\rlap{\lower 3.5 pt \hbox{$\mathchar \sim$}} \raise 1pt \hbox {$<$}}
\def\rsa{\rlap{\lower 3.5 pt \hbox{$\mathchar \sim$}} \raise 1pt \hbox {$>$}}

\def\msbar{\ifmmode{\overline{\rm MS}} \else{$\overline{\rm MS}$} \fi}
\def\drbar{\ifmmode{\overline{\rm DR}} \else{$\overline{\rm DR}$} \fi}

\begin{document}

\hfill\vbox{\baselineskip14pt
            \hbox{UT-647}
            \hbox{KEK-TH-363}
            \hbox{KEK Preprint 93-53}
            \hbox{June 1993}}
\vspace{10mm}

\baselineskip22pt
\begin{center}
\Large	Two-loop renormalization of gaugino mass in
supersymmetric gauge model\footnote{Work supported in part by
Soryushi Shogakukai}
\end{center}
\vspace{10mm}

\begin{center}
\large 	Youichi~Yamada
\end{center}
\vspace{0mm}

\begin{center}
\begin{tabular}{c}
Department of Physics, University of Tokyo, Bunkyo-ku, Tokyo 113, Japan\\
\\
and\\
\\
Theory Group, KEK, Tsukuba, Ibaraki 305, Japan\\
\end{tabular}
\end{center}

\vspace{20mm}
\begin{center}
\large Abstract
\end{center}
\begin{center}
\begin{minipage}{13cm}
\baselineskip=22pt
\noindent
We study the two-loop renormalization group equation
for the running gaugino mass in the supersymmetric
gauge theory. We find that both in the \msbar and \drbar renormalization
schemes, the well-known proportionality of the runnings of the
gaugino mass and the gauge coupling constant is violated
at the two-loop order, even in models with vector supermultiplets only.
Scheme dependence of the running gaugino mass is clarified.
\end{minipage}
\end{center}
\vfill
\newpage

\baselineskip=22pt
\normalsize
\newpage
\normalsize
It has recently been found that the experimentally measured values
of three gauge couplings in the standard model are consistent with the
prediction of the supersymmetric (SUSY) SU(5) grand unified theory
(GUT) \cite{mm}.
The three running gauge coupling constants
$\alpha_i(\mu)=g_i(\mu)^2/(4\pi)(i=1,2,3)$
are found to be unified
at the GUT unification scale $m_U\sim 10^{16}$GeV. One of the
interesting predictions
of this model is that the three running gaugino masses
$m_i(\mu)(i=1,2,3)$ should also be
unified at the same GUT scale and that
the following simple relation
\beq
m_3(\mu)/\alpha_3(\mu)=m_2(\mu)/\alpha_2(\mu)=m_1(\mu)/\alpha_1(\mu)
=m_{1/2}/\alpha_U \label{eq3}
\eeq
holds in the leading order \cite{1loopm}.
Eq.(\ref{eq3}) follows from the relations at $m_U$
\bea
m_3(m_U)=m_2(m_U)=m_1(m_U) &\equiv & m_{1/2} \nonumber \\
\alpha_3(m_U)=\alpha_2(m_U)=\alpha_1(m_U) &\equiv & \alpha_U \label{eq2}
\eea
and the one-loop renormalization group equation \cite{1loopm}
\beq
\frac{d}{dt}\left(\frac{m_i}{\alpha_i}\right)=0,\;\;\;\;\;\;\;\;
t\equiv \ln \mu ,\;\;\;\;\;\;\; i=1,2,3.  \label{eq1}
\eeq
It is worth noting that the above identities hold independent
of any other parameters, such as a particle
content, of the model.

The unification of the gauge coupling constants in the SUSY GUT has
already been studied in the two-loop order.
Just like the three gauge couplings have been measured accurately at the
present $e^+e^-$ colliders, we expect that the next generation
of $pp$ and $e^+e^-$ super-colliders will allow us to measure
the gaugino masses accurately enough \cite{jlc} to test the
unification condition.
It is hence a natural question whether eq.(\ref{eq1}) and, consequently,
the relation (\ref{eq3}) is valid
beyond the one-loop order. It has already been known that in
the two-loop order,
soft SUSY breaking trilinear scalar couplings appear in the renormalization
group equation for gaugino masses \cite{2loopma} and hence
the relation (\ref{eq1}) is violated.
However, it has not been made clear whether the breakdown of the relation
(\ref{eq3}) is solely due to the presence of the SUSY breaking in
the scalar sector or if it is a more general consequence of the
two-loop renormalization group equation.
The relation (\ref{eq1}) is so simple that
one might be tempted to expect that
this relation reflects some symmetry in the gauge interaction sector,
and thus it may hold in all orders.
In this paper we show that the relation (\ref{eq1}) is violated
at the two-loop order even in the gauge interaction sector.
The proportionality of the running gaugino masses and the running gauge
coupling constants is hence an effect specific to
the one-loop approximation.

We work in the simplest SUSY gauge model which contains
only the vector supermultiplets. In this model,
the renormalization group equations for $g$ and $m$ are
diagonal with respect to simple gauge subgroups.
We can hence restrict our study to the case where the gauge group $G$
is simple. In this case,
the two-loop renormalization group equations take the following forms:
\beq
\frac{d}{dt}g=\frac{b_1}{(4\pi)^2}g^3+
\frac{b_2}{(4\pi)^4}g^5, \label{eq4}
\eeq
and
\beq
\frac{d}{dt}m=\frac{b^{(m)}_1}{(4\pi)^2}g^2 m +
\frac{b^{(m)}_2}{(4\pi)^4}g^4 m. \label{eq5}
\eeq
Among the coefficients $b$'s and $b^{(m)}$'s, the followings are
already known:
\bea
& b_1=-3C(G) \cite{1loopg},\;\;\;\;\; b_2=-6C(G)^2 \cite{2loopg},
\;\;\;\;\;
b^{(m)}_1=-6C(G)=2b_1 \cite{1loopm},
& \label{eq6} \\
& \displaystyle C(G)\delta^{ab}\equiv \sum_{c,d} f^{acd}f^{bcd}, & \nonumber
\eea
where $f^{abc}$ is the structure constant of $G$.

In the two-loop order, we can express the ratio $m(\mu)/m(m_U)$ in terms of
the running gauge coupling $\alpha$ by integrating (\ref{eq4})
and (\ref{eq5}).
The result is
\beq
\frac{m(\mu)}{m(m_U)}=\frac{\alpha(\mu)}{\alpha(m_U)}\left(
\frac{b_1+b_2 \alpha(\mu)/(4\pi)}{b_1+b_2 \alpha(m_U)/(4\pi)}
\right)^{(b^{(m)}_2/2b_2-1)}. \label{eq7}
\eeq
In the derivation of eq.(\ref{eq7}), we have used the relation
$b^{(m)}_1=2b_1$.
The first factor on the right hand side of (\ref{eq7}) represents the
one-loop relation (\ref{eq1}) while the second factor
represents the two-loop correction. From eq.(\ref{eq7}),
it is easy to see that the relation (\ref{eq1}) holds
at the two-loop order if $b^{(m)}_2=2b_2$ is satisfied.

Strictly speaking, the condition $b^{(m)}_2=2b_2$ is somewhat ambiguous
since the coefficient $b^{(m)}_2$ is renormalization scheme dependent
while $b_2$ is
independent. In this paper, we examine the following two schemes:
the \msbar scheme (dimensional regularization \cite{ms} with
modified minimal subtraction \cite{msbar}) and
the \drbar scheme (dimensional reduction \cite{dr} with
modified minimal subtraction).
In these schemes,
we obtain
\beq
b^{(m)}_2 = \left\{
\begin{array}{ll}
-32C(G)^2 & (\msbar ), \\
-24C(G)^2 & (\drbar ),
\end{array}
\right. \label{eq8}
\eeq
by evaluating the two-loop diagrams of Fig.1 in the Wess-Zumino gauge.
These results are independent of the gauge fixing parameter.
We find a posteriori that the above result in the \msbar scheme
has been already, though implicitly,
given in ref.\cite{2loopm} which studied the two-loop running
mass parameters in non-SUSY models, because
in the Wess-Zumino gauge, the SUSY gauge model is
equivalent to the non-SUSY gauge theory containing
a Majorana fermion in the adjoint gauge representation.
Our result agrees with theirs.

The physical equivalence of two different values of $b^{(m)}_2$
in eq.(\ref{eq8}) can be proved
by using the one-loop relation between the renormalized
gauge couplings in both schemes \cite{drms},
\beq
g(\mu)_{\drbar} = g(\mu)_{\msbar}\left(
1+\frac{C(G)\alpha(\mu)_{\msbar}}{24\pi}\right) , \label{eq9}
\eeq
and that between the running mass parameters
\beq
m(\mu)_{\drbar} = m(\mu)_{\msbar}\left(
1-\frac{C(G)\alpha(\mu)_{\msbar}}{4\pi} \right) . \label{eq10}
\eeq
The above relation is obtained by evaluating the pole mass of the
gaugino in both schemes,
\bea
m_{pole}&=& m(\mu)_{\msbar}\left[ 1+\frac{C(G)\alpha(\mu)_{\msbar}}{4\pi}
         \left( 3\ln\frac{\mu^2}{m(\mu)^2_{\msbar}}+4 \right) \right]
           \nonumber \\
        &=& m(\mu)_{\drbar}\left[ 1+\frac{C(G)\alpha(\mu)_{\drbar}}{4\pi}
         \left( 3\ln\frac{\mu^2}{m(\mu)^2_{\drbar}}+5 \right) \right] .
\eea

By comparing (\ref{eq6}) and (\ref{eq8}), one can easily see that
the condition $b_2^{(m)}=2b_2$ and, consequently,
the relation (\ref{eq1}) are no longer satisfied
at the two-loop order, in the two commonly used renormalization schemes.
Violation of eq.(\ref{eq1}) at the two-loop level is hence a
general property of these models.

The two-loop corrections to the relation (\ref{eq3}) are of the order of
$\alpha_i$. They consist a part of the next-to-leading order
corrections to the one-loop
predictions of the SUSY particle masses \cite{lahanases}
in GUT models with softly broken supersymmetry.
This correction, as well as the threshold corrections at both
the GUT scale \cite{hisano} and the weak scale, will become
important in the future study of
the unification condition
of the gaugino masses when they
are experimentally measured with a sufficient precision \cite{jlc}.

We have studied the two-loop renormalization group equation for the gaugino
mass and found that in the two commonly used renormalization schemes,
\msbar and \drbar, the one-loop
simple relation (\ref{eq1}) between the gaugino mass and the gauge
coupling constant is violated at the two-loop order.
A more complete analysis, including the contribution of chiral
supermultiplets, Yukawa couplings and soft SUSY breaking trilinear
scalar couplings, is needed to study the unification of the
gaugino masses in
realistic models. We will report this analysis elsewhere.

\section*{\large Acknowledgements}
We would like to thank K. Hagiwara, H. Murayama, Y. Okada and A. Yamada
for useful help and discussions. We also thank Soryushi Shogakukai for
financial support.


\def\PL #1 #2 #3 {Phys.~Lett. {\bf#1} (#3) #2 }
\def\NP #1 #2 #3 {Nucl.~Phys. {\bf#1} (#3) #2 }
\def\ZP #1 #2 #3 {Z.~Phys. {\bf#1} (#3) #2 }
\def\PR #1 #2 #3 {Phys.~Rev. {\bf#1} (#3) #2 }
\def\PP #1 #2 #3 {Phys.~Rep. {\bf#1} (#3) #2 }
\def\PRL #1 #2 #3 {Phys.~Rev.~Lett. {\bf#1} (#3) #2 }
\def\PTP #1 #2 #3 {Prog.~Theor.~Phys. {\bf#1} (#3) #2 }
\def\ib #1 #2 #3 {{\it ibid.} {\bf#1} (#3) #2 }
\def\etal {{\it et al}.}
\def\eg {{\it e.g}.}
\def\ie {{\it i.e}.}


\vspace{20mm}
\section*{Figure Captions}
\renewcommand{\labelenumi}{Fig.\arabic{enumi}}
\begin{enumerate}

\vspace{6mm}
\item
Two-loop diagrams which contribute to the gaugino mass renormalization
in a model with vector supermultiplet only.

\end{enumerate}

\end{document}